\documentclass[showpacs,aps,pra,floats,twocolumn]{revtex4-1}

\pdfoutput=1
\usepackage{graphicx}
\usepackage{float}
\usepackage{braket}
\usepackage{color}
\usepackage{mathtools}
\usepackage{amsmath}
\usepackage{amssymb}
\usepackage{hyperref}
\usepackage{float}
\usepackage{epstopdf}

\begin{document}

\newcommand{\sigp}{\sigma^+}
\newcommand{\sigm}{\sigma^-}
\newcommand{\Gspph}{\Gamma^{\sigma^+}_0}
\newcommand{\Gsmph}{\Gamma^{\sigma^-}_0}
\newcommand{\Gcdph}{\Gamma^\mathrm{cd}_0}
\newcommand{\vc}[1]{{\boldsymbol{\mathrm{#1}}}}
\newcommand{\comment}[1]{}
\newcommand{\remove}[1]{}
\newcommand{\quot}[1]{\textquotedblleft{}#1\textquotedblright}
\newcommand{\un}{\mathrm}
\newcommand{\B}{\langle B \rangle}
\newcommand{\blue}[1]{{\color{blue}#1}}
\newcommand{\red}[1]{{\color{red}#1}}
\makeatletter
\newcommand{\vast}{\bBigg@{4}}
\newcommand{\Vast}{\bBigg@{5}}
\makeatother
\title{Spectral asymmetries in the resonance fluorescence of two-level systems under pulsed excitation}
\author{Chris Gustin}
\email{c.gustin@queensu.ca}
\author{Ross Manson}
\author{Stephen Hughes}
\affiliation{\hspace{-40pt}Department of Physics, Engineering Physics, and Astronomy, Queen's University, Kingston, Ontario K7L 3N6, Canada\hspace{-40pt}}
\date{\today}

%



\begin{abstract}
We present an open-system master equation  study of the coherent and incoherent resonance fluorescence spectrum from a two-level quantum system under coherent pulsed excitation. Several pronounced features which differ from the fluorescence under a constant drive are highlighted, including a multi-peak structure and a pronounced off-resonant spectral asymmetry, in stark contrast to the conventional symmetrical Mollow triplet. We also study semiconductor quantum dot systems using a polaron master equation, and show how the key features of dynamic resonance fluorescence change with electron--acoustic-phonon coupling.
\end{abstract}



\maketitle

The theory of resonant scattering of light from a two-level system (TLS) is a major achievement in quantum optics and provides an experimentally accessible gateway to probing strong-field quantum optics. In recent decades, advances in the ability to coherently manipulate atomic systems with light has allowed for a breadth of technological innovations which harness the quantum mechanical properties of these systems~\cite{dowling14}. Furthermore, quantum dots (QDs) -- semiconductor materials confined in three dimensions, with excited electron-hole pairs (excitons) mimicking the behaviour of an excited atom, can serve as  "artificial atoms", maintaining the physics of the quantized system's interaction with the electromagnetic field, but with tunable properties and potential for scalability~\cite{lodahl15}. Semiconductor QDs have been the subject of much recent research for their potential as sources of quantum light, particularly single and entangled photons ~\cite{buckley12}. While constant excitation with a continuous wave (cw) laser drive can be used to create a TLS single-photon source, often technological proposals require a deterministic source -- one that can be triggered on-demand. This is typically done by  an optical pulse, which  renders resonance fluorescence (RF) of a TLS a genuine time-dependent quantum dynamical process.

The usual features of the RF spectrum under strong cw excitation manifest as the so-called Mollow triplet~\cite{mollow69}, where the power spectrum of the scattered field takes on a characteristic three-peak resonance structure due to radiative transitions between eigenstates of the system Hamiltonian, as well as a delta function peak at the (monochromatic) drive frequency corresponding to coherent elastic scattering. However, under excitation by a short pulse, the RF spectrum can take on features which obscure or eliminate this characteristic spectrum, especially under off-resonant excitation. The pulsed RF spectrum has been studied theoretically in atomic systems~\cite{cavalieri87,edwards82,rza84,flor84,buffa88,ho88}, and more recently in QD-cavity systems for on-resonance excitation~\cite{moelbjerg12}, where a dynamic spectrum has been observed in the presence of cavity coupling~\cite{fischer16}.

In this Letter, we describe the unique  features of pulsed RF spectra in  depth using a master equation approach, and explore the different effects under time-dependent excitation, which are of interest to emerging experimental studies of pulsed quantum optical systems. In particular, we separate both the coherent and incoherent spectra under pulsed excitation of a single TLS incorporating the dissipative processes of spontaneous emission and pure dephasing. Using an off-resonant drive, we describe how a pronounced spectral asymmetry can arise under different conditions, and explain its origin in terms of the rate of optical transitions between dressed eigenstates. We investigate the different regimes under which a multi-peak structure arises, and describe the role of the excited state in the off-resonant spectrum. In semiconductor QD systems, acoustic-phonon--electron scattering  alters the physics of two-level quantum optics by introducing further decoherence, including non-Markovian bath coupling~\cite{nazir16,ramsay10,krummheuer02,besombes01,vagov02,forstner03}; with pulsed excitation, we study this fundamental quantum interaction in detail, explaining its rich features, then discuss the main effects of including phonon coupling via a polaron master equation technique, which  rigorously describes the dynamics of optically pulsed QD systems ~\cite{ross16, gustin17,fischer15}.

The system Hamiltonian $H_S$ of a TLS, under laser excitation in the dipole approximation, is   %
\begin{equation}\label{one}
H_S(t) = \Delta\sigma^+\sigma^- + \frac{\Omega (t)}{2}(\sigma^+ + \sigma^-),
\end{equation}
in a frame rotating at the frequency of the laser $\omega_L$,
where $\sigma^+$, $\sigma^-$, are the Pauli  operators between the ground and excited states, 
$\Delta \equiv \omega_e - \omega_L$ is the detuning between the excited state and laser frequency, and $\Omega (t)$ is the Rabi frequency. To derive the time-dependent dynamics of the system incorporating dissipation, we employ a Lindblad master equation for the system density operator $\rho$:
\begin{equation}
\label{two}
\frac{d}{dt}\rho = -\frac{i}{\hbar}[H_S(t),\rho] + \frac{\gamma}{2}\mathcal{L}[\sigma^-]\rho + \frac{\gamma '}{2}\mathcal{L}[\sigma^+\sigma^-]\rho,
\end{equation}
where the Lindblad terms ($\mathcal{L}[A]\rho = 2A\rho A^{\dagger} -A^{\dagger} A\rho - \rho A^{\dagger}A$) represent spontaneous emission with rate $\gamma$ and a pure dephasing  with rate $\gamma '$. For a TLS, the total spectrum $S(\omega)$  is~\cite{cui06}
\begin{equation}\label{spect}
S(\omega) = \text{Re}\left[\!\int_{0}^{\infty}\!dt\!\int_0^{\infty}\!d\tau \langle \sigma^+(t)\sigma^-(t+\tau)\rangle     e^{i(\omega-\omega_L) \tau}         \right],
\end{equation}
where  $\langle \sigma^+(t)\sigma^-(t+\tau)\rangle$ is calculated from the quantum regression theorem.
The incoherent spectrum is $S_{\text{inc}}(\omega) = S(\omega) - S_{\text{coh}}(\omega)$, where the coherent spectrum is
%
\noindent $
S_{\text{coh}}(\omega) = \text{Re}\left[\!\int_{0}^{\infty}\!dt\!\int_0^{\infty}\!d\tau \langle \sigma^+(t)\rangle \langle \sigma^-(t+\tau)\rangle     e^{i(\omega-\omega_L) \tau}         \right]$.

It is useful to recognize that for slowly varying pulses (the adiabatic limit), the time-dependent eigenstates of Eq.~\eqref{one} give the dressed states of the system at a given time $t$. These states, $\ket{\pm}$, have energies $ \epsilon_{\pm} = \frac{\Delta}{2} \pm \frac{\Omega_R}{2}$, where $\Omega_R \equiv \sqrt{\Omega^2+\Delta^2}$ is the effective Rabi frequency, and 
$\ket{\pm} = \frac{1}{\sqrt{1+\kappa_{\pm}^2}} (\ket{g}+\kappa_{\pm}\ket{e})$
with $\kappa_{\pm} = {\Omega}/{[\pm \Omega_R - \Delta]}$. In a fully quantum-mechanical model where the electric field is quantized, 
 the states $\ket{\pm}$ form a manifold $\ket{\pm,N}$ denoted by photon number $N$. The optical transitions $\ket{+,N}\! \rightarrow\! \ket{-,N-1}$ (which we denote as a $+\! \rightarrow\! -$ transition) and $\ket{-,N}\! \rightarrow\! \ket{+,N-1}$ ($-\! \rightarrow\! +$), along with the $\ket{+,N}\! \rightarrow\! \ket{+,N-1}$ and $\ket{-,N}\! \rightarrow\! \ket{-,N-1}$ create the well-known Mollow triplet. Note that when $\Delta =0$, the dressed states are equal mixtures of the ground and excited states. For nonzero detuning, however, in the limit of a weak drive, $\ket{-}$ tends to the ground state, and $\ket{+} $ tends to the excited state.
\begin{figure}[h]
        \centering
        \includegraphics[width=1\linewidth]{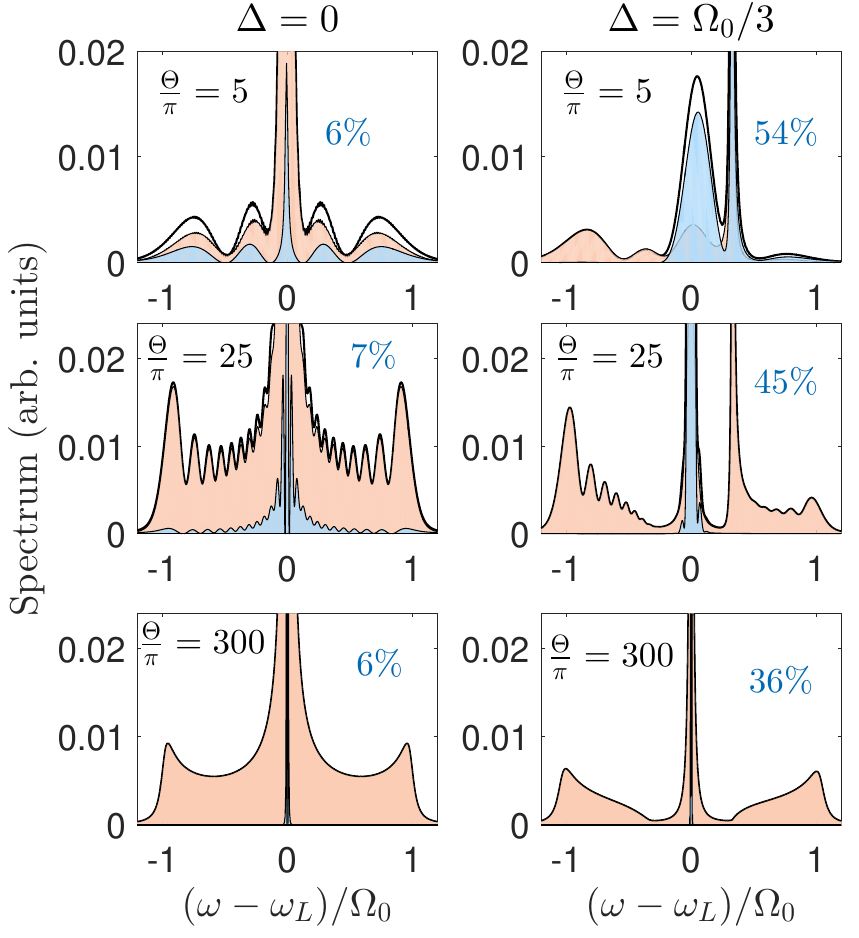}
        \vspace{-0.2cm}
        \caption{\small Total (black line), incoherent (orange), and coherent (blue) spectra for a pulsed TLS with $\gamma = \Omega_0/40$, $\gamma' = 0$, for varying pulse widths. Each row of spectra (where the left is for a resonant laser, and the right a laser detuned by $\Delta$) is normalized to the maximum amplitude of the total spectra of the left (resonant driving) case. Also shown in blue for each case is the percent of coherently scattered light (integrated spectral intensity) to the total spectral intensity.}
        \label{dynamicmollow}
\end{figure}

In Fig.~\ref{dynamicmollow}, we show the incoherent, coherent, and total spectra for a two-level system driven at a maximum Rabi frequency $\Omega_0$ for varying pulse widths. The pulse width is expressed in terms of the dimensionless pulse area, $\Theta = \int_{-\infty}^{\infty}\Omega(t)dt$, where we have chosen a Gaussian pulse of the form $\Omega(t) = \Omega_0 \exp{[-\pi(\frac{\Omega_0 t}{\Theta})^2]}$. At low pulse areas, both the incoherent and coherent spectra take on a characteristic multi-peak structure, due to interference effects. On resonance, an analytical criterion for the emergence of these sidepeaks has been derived by Moelbjerg \emph{et al.}~\cite{moelbjerg12}, where the solutions $t_n$ to the equation $\int_{-t_n}^{t_n}\Omega(t')dt' - 2\Omega(t_n)t_n = (2n + \frac{1}{2})\pi$ give the times at which the scattered field interferes constructively at frequency $\Omega(t_n)$, leading to sidepeaks in the spectrum at locations $\omega - \omega_L = \pm\Omega(t_n)$. As the pulse area increases, these peaks merge into a continuum, which levels off at the locations of the cw Mollow sidepeaks at $\pm \Omega_0$. To see only dynamical features in the spectrum, the pulse width should be shorter than the lifetime of the excited state, such that  $\Theta \ll \frac{\Omega_0}{\gamma}$, or for Fig.~\ref{dynamicmollow}, $\frac{\Theta}{\pi} \ll 13$. With off resonance driving (not studied in~\cite{moelbjerg12}), similar behaviour is observed, but with an additional peak appearing in the spectra for short pulse widths, corresponding to the excited state energy level. For long pulses, the dressed states of the system have time to fully develop, and this peak disappears. Note that changing the sign of the detuning $\Delta$ simply reflects each spectra about the vertical axis.

Off-resonance, a significant asymmetry is seen in the incoherent spectra. The incoherent spectra arises from transitions between the field-dressed states of the system, and this asymmetry can be explained in terms of adiabatic evolution of the system eigenstates over time~\cite{edwards82, cavalieri87}. Since the system starts in the ground state, for non-zero detunings, it will adiabatically follow  $\ket{-}$, provided the system evolution is slow. More precisely, the criterion for pulsed adiabatic evolution  is~\cite{tannor}
\begin{equation}\label{adia}
\Bigg|\frac{2\Delta}{\tau^2} \frac{\Omega(t)t}{([\Omega(t)]^2+\Delta^2)^{3/2}}\Bigg| \ll 1,
\end{equation}
where $\tau = \Theta/(\sqrt{\pi}\Omega_0)$ is the 1/e pulse half width. Eq.~\eqref{adia} can be satisfied for, e.g.,  long pulse widths and/or large detunings. For rapidly-varying pulse envelopes (e.g., a square pulse), this asymmetry is flipped. Note that this asymmetry vanishes as the off-resonant pulse width is increased; for long pulse widths, the two-level system absorbs and emits multiple photons, and the photon statistics tend towards a Poissonian distribution~\cite{fischer17}. In this limit, the population of the system eigenstates are determined by radiative transitions from higher-lying manifolds in the dressed-state picture, and thus the system obeys the principle of detailed balance. Explicitly, if a photon is emitted from the $\ket{+,N}$ state, transitioning the TLS into state  $\ket{-,N-1}$  (a $+\! \rightarrow\! -$ transition), the TLS can only reach state $\ket{+,N-1-n}$  ($n \in \mathbb{Z}$) by undergoing a $-\! \rightarrow\! +$ transition, and vice-versa. 
%
\begin{figure}
        \centering
        \includegraphics[width=1\linewidth]{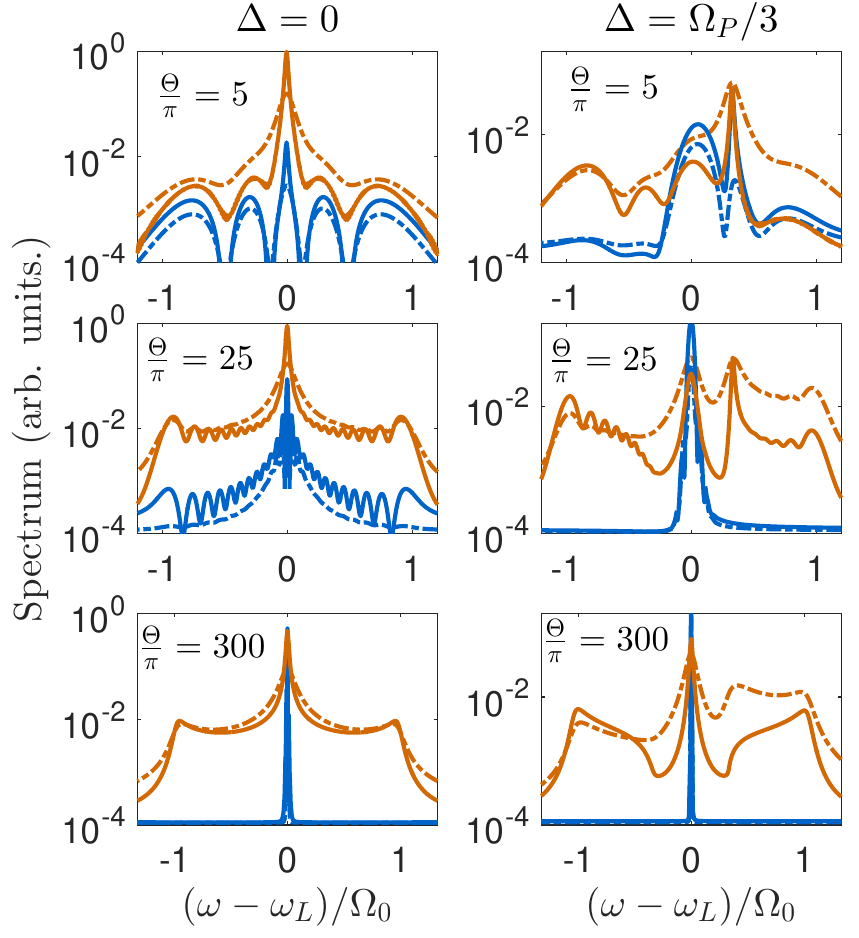}
         \vspace{-0.2cm}
        \caption{\small Spectra from Fig.~\ref{dynamicmollow} on a semilog scale, but with incoherent (orange) and coherent (blue) spectra also shown for a pure dephasing $\gamma' = \Omega_0/10$ (dash-dotted lines).}
        \label{dynamicdeph}
\end{figure}

The principle of detailed balance, however, only strictly applies in the absence of pure (non-radiative) dephasing. In Fig.~\ref{dynamicdeph}, we plot the pulsed RF spectra under the same conditions as in Fig.~\ref{dynamicmollow}, but with the incoherent and coherent spectra in the presence of additional pure dephasing also included. We also use a semilog scale in this case, to better reveal the amplitude of some of the larger peaks. With dephasing, a clear asymmetry is also seen on the blue side of the spectrum, with the $+\! \rightarrow\! -$ transitions larger. To understand the physics behind this asymmetry, it is instructive to consider the cw limit, where exact analytical solutions for the spectra are known. The spectral weight (area) of a peak caused by radiative transitions (as with the incoherent spectrum) is proportional to the product of the population of the initial state (pre-transition) and the transition rate~\cite{florescu04}. Explicitly, the spectral weight $\Gamma$ of a transition with rate $\tilde{\gamma}$ is 
\begin{equation}
\Gamma_{\substack{+-\\ -+}} = \tilde{\gamma} \text{tr}\big[\ket{\pm}\bra{\pm} \rho(\infty)\big]\ \big|\!\bra{\mp}\sigma^-\ket{\pm}\!\big|^2.
\end{equation}
The transition rate $\tilde{\gamma}$ is not exactly solvable analytically for an off-resonantly driven Mollow triplet, but it can be shown by considering the determinant of the characteristic equation of the optical Bloch equations in matrix form that this rate is the same for both $\ket{+}$ and $\ket{-}$ states. The relative spectral weight of these transitions gives the magnitude of this asymmetry:
\begin{align}
\frac{\Gamma_{+-}}{\Gamma_{-+}}  &= \frac{\kappa_+^2(1+\kappa_-^2)}{\kappa_-^2(1+\kappa_+^2)} \nonumber \\ &  \times 
\left[\frac{1+(\kappa_+^2-1)\langle \sigma^+ \sigma^-\rangle + 2\kappa_+ \text{Re}\{\langle \sigma^- \rangle \}}{1+(\kappa_-^2-1)\langle \sigma^+ \sigma^-\rangle + 2\kappa_- \text{Re}\{\langle \sigma^- \rangle \}}\right]
\end{align}
where the  terms~${\langle \sigma^- \rangle = -\frac{i \Omega}{2}\Big(\frac{\gamma_p + i\Delta}{\gamma_p^2+\Delta^2 + \Omega^2\,{\gamma_p}/{\gamma}}\Big)}$ and ${\langle \sigma^+ \sigma^- \rangle = \frac{1}{2}\Big[1+\frac{\gamma}{\gamma_p}\big(\frac{\gamma_p^2+\Delta^2}{\Omega^2}\big)\Big]^{-1}}$ are found from the steady state solution of Eq.~(\ref{two}) for a cw Rabi frequency $\Omega$, with $\gamma_p = \frac{1}{2}(\gamma+\gamma')$~\cite{ulhaq13}. Physically, this asymmetry is due to the fact that pure dephasing acts to reduce the bare-state basis polarization. In the dressed state picture, this asymmetry appears as non-radiative transitions between eigenstates, which violates the principle of detailed balance. Since, off resonance, the $\ket{+}$ state is more like the excited TLS state, it has a higher radiative transition rate, which is balanced in the case of no dephasing by the lower steady state $\ket{+}$ population. Dephasing increases this state's population, and thus spectral weight. The physics of the cw case can be qualitatively extrapolated back to the pulsed case by considering the time-dependent eigenstates.
In the application of our results to realistic semiconductor QD systems, it is essential to incorporate into the model the effects of longitudinal acoustic (LA) phonon scattering. The total Hamiltonian (system + reservoir) in this case becomes
\begin{align}
H_{\text{tot}}(t) &= \Delta\sigma^+\sigma^- + \frac{\Omega (t)}{2}(\sigma^+ + \sigma^-) 
\nonumber \\ + & 
\sum\limits_{\mathbf{q}}\omega_{\mathbf{q}}b_{\mathbf{q}}^{\dagger}b_{\mathbf{q}}+ \sigma^+\sigma^-\sum\limits_{\mathbf{q}}\lambda_{\mathbf{q}}(b_{\mathbf{q}}^{\dagger}+b_{\mathbf{q}}),
\end{align}
where $b_{\mathbf{q}}$ ($b_{\mathbf{q}}^{\dagger}$) are annihilation (creation) operators corresponding to phonon modes for wavevector $\mathbf{q}$, coupled to the exciton by coupling constants $\lambda_{\mathbf{q}}$. By treating the phonon modes as a reservoir and applying a unitary "polaron" transformation, we derive a master equation which captures nonperturbative electron-phonon coupling effects and then treats the pulse as a perturbation about this polaron frame:
\begin{align}
\label{me}
\frac{d}{dt}\rho &= -i[H_S'(t), \rho] + \frac{\gamma}{2}\mathcal{L}[\sigma^-]\rho
\nonumber \\  
&-\int_0^{\infty}d\tau \sum\limits_{\mathclap{m = \{g, u\}}}  (G_m(\tau) 
\nonumber \\   
& \times [X_m(t),e^{-iH_s(t)\tau}X_m(t)e^{iH_s(t)\tau} \rho] + \text{H.c.}, 
\end{align}
which is similar to Eq.~\eqref{two} for $\gamma' = 0$ with an additional term corresponding to incoherent electron-phonon scattering (see~\cite{ross16} for details). The phonon interaction is characterized by the phonon spectral function $J(\omega) = \alpha \omega^3 \exp{\left[\!-\!\frac{\omega^2}{2\omega_b^2}\right]}$, where we choose a coupling strength of $\alpha = 0.06 \ \text{ps}^2$  and a phonon bath frequency cut-off of $\omega_b = 1 \ \text{meV}$, both consistent with the experimental-theory results of Weiler \emph{et al.}~\cite{weiler12}. The polaron transform renormalizes the system Hamiltonian by coherently reducing the effective drive strength $\Omega(t)$ and introducing a Lamb-type shift in the exciton resonant frequency, but we have assumed that these effects can be absorbed into the original definitions of these parameters and neglected henceforth (such that $H_S' = H_S$). Note that to calculate the dot-emitted spectrum in the laboratory frame in terms of the polaron-transformed operators, one must multiply the two-time correlation function in Eq.~\eqref{spect} by an additional factor (related to the polaron Green functions)~\cite{roy12}).
This factor captures non-Markovian lattice relaxation effects, leading to a broad phonon sideband which is asymmetric at low temperatures superimposed over the incoherent spectrum, and also reduces the coherent spectrum amplitude~\cite{roy12,ilessmith17}. However, it is usually the case in semiconductor experiments that the QD is weakly coupled to a cavity and the output emission is through the cavity mode. In this case, the field operators $a \ (a^{\dagger})$ (which are unaffected by the polaron transform) describing the cavity mode can be shown to be proportional to the exciton operators $\sigma^- (\sigma^+)$~\cite{hargart16} and the additional phonon factor need not be included to calculate the cavity-emitted spectrum.

For our semiconductor calculations, we employ QD-appropriate physical units, choosing $\Omega_0 = 1$ meV, $\gamma' = 0$, $\gamma = 10 \ \mu \text{eV}$ (corresponding to a Purcell factor of around 10), temperature $T = 4$ K, and $\Theta = 5\pi$ (pulse FWHM of 9.7 ps).
\begin{figure}[ht]
        \centering
        \includegraphics[width=1\linewidth]{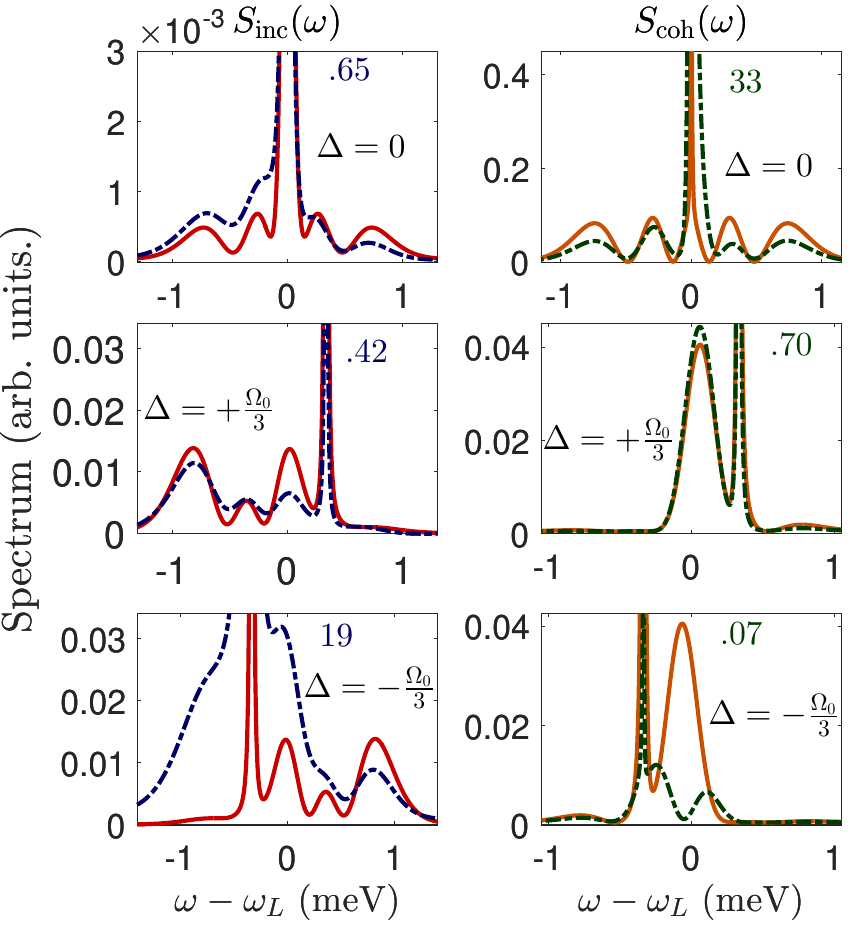}
        \vspace{-0.2cm}
        \caption{\small Incoherent (left) and coherent (right) RF spectra for a driven QD without the exciton-phonon interaction (solid red and orange) and with exciton-phonon coupling where the spectrum is emitted from a weakly-coupled cavity (dashed-dotted; dark blue and green). The exciton-laser detuning is (top) $\Delta = 0$, (middle) $\Delta = 0.33$ meV, (bottom) $\Delta = -0.33$ meV. The incoherent and coherent spectra are each normalized (separately) to the maximum amplitude in the spectra without phonons. Also shown is the maximum amplitude of the off-screen peak of each spectrum with phonon-coupling.}
        \label{phonon}
\end{figure}
In Fig.~\ref{phonon} we present spectra with and without phonon coupling for a resonant drive, as well as for drives red and blue detuned with respect to the exciton. In contrast to previous studies on the RF spectra of QDs under pulsed excitation which only considered electron-LA-phonon scattering~ via a non-polaronic time-dependent dephasing~\cite{moelbjerg12}, our full phonon model finds a notable asymmetry even in the cavity-emitted spectrum \emph{on resonance}, due to phonon emission-assisted radiative decay, which is more probable than phonon absorption at low temperatures~\cite{roy12}. Symmetry is restored in the cw case~\cite{roy12,ulhaq13} as the principle of detailed balance dictates the spectral weights when multi-photon emission dominates. For the same reason, the spectra are heavily modified for negative detunings; the phonon interaction now greatly increases the amplitude of the exciton peak in the incoherent spectrum, and decreases (relative to the amplitude of the peak when the detuning is positive) it in the coherent spectrum. This is because the exciton is populated via incoherent phonon emission, in contrast to the coherent no-phonon case of quasiresonant excitation by the continuum of Rabi frequencies contained in the pulse. Furthermore, phonons completely change the multi-peak spectra around the laser frequency for this detuning -- removing the center peak entirely for the coherent spectra. Consistent with~\cite{moelbjerg12}, we find that strong phonon coupling renders the multi-peak structure more difficult to resolve, but does not completely obscure it even with our large phonon coupling strength.

To conclude, we have presented a theoretical and computational study of pulsed RF spectra of two-level systems under on and off resonance excitation, described how these can differ from RF spectra under a constant drive, and extended our general results to semiconductor QD systems via a rigorous polaron transform model of electron-phonon coupling. We have elucidated the different ways spectral asymmetry can arise by considering the adiabatic evolution of the dressed states, non-radiative pure dephasing, and phonon-assisted transitions.
\section*{Funding Information}
Natural Sciences and Engineering Research Council of Canada (NSERC). 
\bibliography{Bib_1005}

\end{document}